\def\kF{k_{\text{F}}}
\def\gso{g_{\text{so}}}
\def\kB{k_{\text{B}}}
\def\NF{N_{\text{F}}\,}
\def\me{m_{\text{e}}}
\def\Hcone{H_{\text{c}1}}
\def\Hctwo{H_{\text{c}2}}
\def\be{\begin{equation}}
\def\ee{\end{equation}}
\def\bea{\begin{eqnarray}}
\def\eea{\end{eqnarray}}
\def\bse{\begin{subequations}}
\def\ese{\end{subequations}}

\documentclass[prl,twocolumn,showpacs,preprintnumbers,amsmath,amssymb]{revtex4}
\usepackage{graphicx}
\usepackage{dcolumn}
\usepackage{bm}
\begin{document}
\title{Anomalous Pinning Fields in Helical Magnets: Screening of the Quasiparticle
       Interaction 
}
\author{T.R. Kirkpatrick$^{1}$ and D. Belitz$^{2}$}
\affiliation{$^{1}$Institute for Physical Science and Technology and Department
                   of Physics, University of Maryland, College Park, MD 20742\\
             $^{2}$Department of Physics, Institute of Theoretical Science, and
                   Materials Science Institute, University of Oregon, Eugene, OR
                   97403}
\date{\today}
\begin{abstract}
The spin-orbit interaction strength $\gso$ in helical magnets determines both
the pitch wave number $q$ and the critical field $\Hcone$ where the helix
aligns with an external magnetic field. Within a standard
Landau-Ginzburg-Wilson (LGW) theory, a determination of $\gso$ in MnSi and FeGe
from these two observables yields values that differ by a factor of $20$. This
discrepancy is remedied by considering the fermionic theory underlying the LGW
theory, and in particular the effects of screening on the effective
electron-electron interaction that results from an exchange of helical
fluctuations.
\end{abstract}


\maketitle

Chiral itinerant ferromagnets such as MnSi \cite{Ishikawa_et_al_1976,
Pfleiderer_et_al_1997} and FeGe \cite{Lebeche_Bernhard_Freltoft_1989} have
recently attracted considerable attention. Both of these systems crystallize in
the cubic B20 structure, which lacks inversion symmetry, and as a result
spin-orbit coupling effects are important for the magnetic properties. They
both exhibit spiral or helimagnetic spin order at low temperatures (below about
28.5 K at ambient pressure in MnSi, and below about 279 K in FeGe,
respectively), which is believed to be generated by a Dzyaloshinski-Moriya term
\cite{Dzyaloshinski_1958, Moriya_1960} in the free energy. The pitch wavelength
of the helix is large compared to a microscopic length scale; this reflects the
weakness of the spin-orbit interaction. The pitch wave vector is $q = 0.035\
\AA^{-1}$ in MnSi \cite{Ishikawa_et_al_1985} and $q \approx 0.009\ \AA^{-1}$ in
FeGe \cite{Lebeche_Bernhard_Freltoft_1989}. In other parts of the phase diagram
in MnSi, striking non-Fermi-liquid behavior has been observed in
low-temperature transport measurements \cite{Pfleiderer_Julian_Lonzarich_2001}.

In an isotropic electron system there would be no preferred direction for the
pitch vector of the helix. In real materials, the underlying crystal lattice
pins the helix. The terms in the free energy that cause this pinning are of
higher order in the spin-orbit interaction, and hence represent an energy scale
that is even weaker than those that lead to the formation of the helix. In
MnSi, the pinning is in the $(1,1,1)$-direction (or equivalent); in FeGe, it is
in the $(1,0,0)$-direction (or equivalent) close to the transition, and in the
$(1,1,1)$-direction at lower temperatures. An external magnetic field makes it
energetically favorable for the helix to align with the field, and this
competes with the crystal-field effects. As a result, upon applying a magnetic
field in, say, the $(0,0,1)$-direction to MnSi in the helical phase, the pitch
vector rotates away from $(1,1,1)$ until it aligns with the field direction at
a critical field strength $H = \Hcone$. This field strength marks the boundary
of the so-called conical phase, which is characterized by a homogeneous
magnetization superimposed on the helix that is aligned with the field. Upon
further increasing the field, the homogenous component of the magnetization
increases, and the amplitude of the helix continuously decreases until it
disappears at a second critical field, $\Hctwo$, where the system enters a
field-polarized phase with a homogeneous magnetization.

The above considerations make it clear that one can obtain a measure of the
spin-orbit interaction strength from measuring either the pitch wave number or
the critical field $\Hcone$. It is a puzzling, but overlooked, fact that
interpreting the results within the existing theoretical framework yields
values for the spin-orbit interaction strength that differ by a factor of about
$25$. It is the purpose of the present Letter to resolve this discrepancy.

In order to frame our discussion of these various effects, let us consider the
Landau-Ginzburg-Wilson (LGW) theory that has been commonly used to describe
helical magnets \cite{Bak_Jensen_1980}. If the phase transition is either
continuous or weakly first order, then the classical behavior of the system
close to the transition can be described by an action
\bse
\label{eqs:1}
\be
S = S_0 + S_{\text c} + S_{\text{cf}} + S_H.
\label{eq:1a}
\ee
Here $S_0$ is the usual action for a classical Heisenberg ferromagnet
\cite{Ma_1976},
\be
S_0 = \int d{\bm x}\ \left[\frac{r_0}{2}\,{\bm M}^2({\bm x}) +
\frac{a}{2}\,\left(\nabla{\bm M}({\bm x})\right)^2 + \frac{u}{4}\,{\bm
M}^4({\bm x})\right].
\label{eq:1b}
\ee
Here ${\bm M}({\bm x})$ is the three-component order parameter whose
expectation value is proportional to the magnetization, $r_0$ is the bare
distance from the critical point, and $a$ and $u$ are parameters that depend on
the microscopic details of the system. $(\nabla{\bm M})^2$ is a shorthand
notation for $\sum_{i,j}\partial_i M_j\,\partial^i M^j$. $S_0$ is invariant
under separate rotations in order-parameter space and real space.

$S_{\text c}$ is the leading chiral term induced by the spin-orbit interaction
\cite{Dzyaloshinski_1958, Moriya_1960},
\be
S_{\text c} = \frac{c}{2}\int d{\bm x}\ {\bm M}({\bm
x})\cdot\left({\bm\nabla}\times{\bm M}({\bm x})\right).
\label{eq:1c}
\ee
The coupling constant $c$ is proportional to the dimensionless spin-orbit
interaction strength $g_{\text{so}}$, and on dimensional grounds we have $c =
a\kF\gso$. Note that $S_{\text{c}}$ is still invariant under joint rotations in
order-parameter space and real space, but not under spatial inversions. This
terms can therefore be present only in systems that are not inversion
invariant. The chiral nature of the curl produces the helical ground state, and
the handedness of the helix depends on the sign of $c$. We assume $c>0$ without
loss of generality.

$S_{\text{cf}}$ is the largest term that describes the crystal-field effects
that couple the magnetization to the underlying lattice. For a cubic lattice,
a representative contribution to $S_{\text{cf}}$ reads \cite{Bak_Jensen_1980}
\be
S_{\text{cf}} = b \int d{\bm x}\ \sum_{i}\left(\frac{\partial M_i}{\partial
x_i}\right)^2.
\label{eq:1d}
\ee
The coupling constant $b$ is quadratic in $\gso$ and given by $b = a'\gso^2$
with $\vert a'\vert\approx a$. (Here and it what follows we ignore factors of
$O(1)$.) $S_{\text{cf}}$ breaks the rotational invariance and is responsible
for pinning the helix. The direction of the pinning depends on the sign of b.

Finally, we have a term that couples an external magnetic field ${\bm H}$ to
the magnetization:
\be
S_H = \int d{\bm x}\ {\bm H}({\bm x})\cdot{\bm M}({\bm x}).
\label{eq:1e}
\ee
\ese

In the absence of both an external field and any coupling to the underlying
lattice, it is easy to see that the Eqs.\ (\ref{eq:1b},\ref{eq:1c}) lead to a
helical ground state:
\be
{\bm M}({\bm x}) = m_1\,\left[{\hat{\bm e}}_1\,\cos({\bm q}\cdot{\bm x}) +
{\hat{\bm e}}_2\,\sin({\bm q}\cdot{\bm x})\right].
\label{eq:2}
\ee
Here the unit vectors ${\hat{\bm e}}_1$, ${\hat{\bm e}}_2$, and ${\hat{\bm q}}
= {\bm q}/\vert{\bm q}\vert$ form a right-handed {\it dreibein}.
The amplitude of the helix is given by $m_1 = \sqrt{-(t-aq^2)/u}$. The pitch
vector ${\bm q}$ points in an arbitrary but fixed direction, and in a
mean-field approximation its modulus is given by $q = c/2a + O(\gso^2)$. $q$ is
small compared to the Fermi wave number $\kF$ by virtue of the smallness of
$\gso$. In MnSi, $\kF \approx 3.6\,\AA^{-1}$ \cite{parameters_footnote}, so
$q/\kF \approx 0.01$. Assuming the same value for $\kF$ in FeGe, we have $q/\kF
\approx 0.0025$. The value of $\gso$, which is equal to $q/\kF$ within a factor
of $2$ \cite{factor_2_footnote}, is thus
\be
\gso \approx \begin{cases} 0.01\quad&\text{(MnSi)}\cr
                           0.0025     &\text{(FeGe).}
                    \end{cases}
\label{eq:3}
\ee

As discussed above, a magnetic field tends to align the helix away from the
pinning direction that is ultimately determined by the spin-orbit interaction,
and hence the magnitude of the field necessary to depin the helix provides
another estimate for $\gso$. In MnSi, the best studied helimagnet, the helix in
zero field is pinned in the $(1,1,1)$-direction, which implies that the
coefficient $b$ in Eq.\ (\ref{eq:1d}) is negative ($b>0$ leads to pinning in
the $(1,0,0$-direction) \cite{Bak_Jensen_1980}. At ambient pressure, and not
too close to the transition temperature, the experimental value for the field
$\Hcone$ defined above, where the pitch vector ${\bm q}$ aligns with the field
direction, is $\Hcone \approx 0.1\,{\text{T}}$ \cite{Muhlbauer_et_al_2009}. In
the same region, the experimental value for the field $\Hctwo$, where the helix
vanishes, varies between $0.4\,\text{T}$ and $0.55\,\text{T}$. Together with
the corresponding experimental results for FeGe
\cite{Lebeche_Bernhard_Freltoft_1989}, we thus have a ratio
\be
\Delta_{\text{exp}} \equiv \Hcone/\Hctwo \approx \begin{cases}  0.2 \quad &
\text{(MnSi)}\cr 0.1 & \text{(FeGe)}. \end{cases}
\label{eq:3}
\ee

Comparing these experimental results with theoretical estimates leads to a
puzzle. The Eqs.\ (\ref{eqs:1}) can be analyzed in detail to yield the critical
fields $\Hcone$ and $\Hctwo$ \cite{Plumer_Walker_1981, us_tbp}, but for our
present purposes the following simple considerations suffice. $\Hcone$ is
roughly determined by the magnetic energy, given by $S_H$, being equal to the
pinning energy, which is given by $S_{\text{cf}}$. $\Hctwo$ is roughly
determined by the magnetic energy being equal to the chiral energy, which is
given by $S_{\text c}$. For a homogeneous magnetic field, the coupling in $S_H$
is to the homogeneous magnetization, $m_0 \equiv \int d{\bm x}\ M({\bm x}) =
\chi H$, with $\chi$ the homogeneous magnetic susceptibility. The magnetic
energy is thus of $O(H^2)$, $S_H = \chi H^2$. In Eq.\ (\ref{eq:1d}), the
gradient squared is on the order of $q^2$, and the magnetization is on the
order of the amplitude of the helix, $m_1$. For $\Hcone$ we thus obtain the
estimate
\bse
\label{eqs:5}
\be
\Hcone \approx \gso\,m_1\,q \sqrt{a/\chi}.
\label{eq:5a}
\ee
Applying an analogous estimate to Eq.\ (\ref{eq:1c}), we obtain
\be
\Hctwo \approx m_1 \sqrt{a\kF\gso\,q/\chi}.
\label{eq:5b}
\ee
\ese
All quantities whose estimates might be questionable thus drop out of the ratio
$\Delta$, and we have the theoretical result from the bare LGW theory
\be
\Delta_{\text{theo}}^{\text{bare}} \approx \sqrt{\gso\,q/\kF} \approx \gso.
\label{eq:6}
\ee
Comparing Eqs.\ (\ref{eq:3}), (\ref{eq:3}), and (\ref{eq:6}), we see that the
experimental values for $\Delta$ are larger than the theoretical expectation by
about a factor of $20$ in MnSi and $40$ in FeGe. We will now show how this
discrepancy can be resolved by considering the screening of the effective
electron-electron interaction that results from the exchange of helical
fluctuations.

We first need to discuss the nature of the dominant fluctuations in a helical
magnet. The helical ground state represents a spontaneous breaking of
translational invariance, and therefore leads to a Goldstone mode or
helimagnon. For the LGW action written above, and as a function of the wave
vector ${\bm k}$ for $\vert{\bm k}\vert \ll q$, the frequency of the helimagnon
reads \cite{Belitz_Kirkpatrick_Rosch_2006a, sign_b_footnote}
\be
\omega_0({\bm k}) = \sqrt{a\,k_{||}^2 + 2\vert b\vert\,{\bm k}_{\perp}^2/3 +
a\,{\bm k}_{\perp}^4/2q^2}.
\label{eq:7}
\ee
Here ${\bm k} = (k_{||},{\bm k})$ has been decomposed into components parallel
and perpendicular to the pitch wave vector ${\bm q}$. We have assumed $b<0$, as
appropriate for MnSi \cite{Bak_Jensen_1980}, and we have neglected corrections
of $O(b) = O(\gso^2)$ to the coefficients of $k_{||}^2$ and ${\bm
k}_{\perp}^4$. For $b=0$, that is, if we neglect all effects of $O(\gso^2)$,
the helimagnon frequency squared lacks a contribution proportional to ${\bm
k}_{\perp}^2$. This is a result of rotational invariance. The crystal-field
term $S_{\text{cf}}$ in the action breaks this invariance, which leads to a
mode that is still soft (because the translational invariance is still broken),
but has a ${\bm k}_{\perp}^2$-term with a small prefactor. It is the
generalization of the well-known magnons in ferromagnets and antiferromagnets
that have a quadratic and linear dispersion relation, respectively.

\begin{figure}[t]
\vskip -0mm
\includegraphics[width=8.0cm]{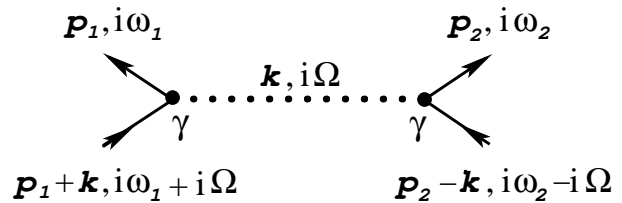}
\caption{The effective quasiparticle interaction due to helimagnons. Note that
the vertices depend on the quasiparticle momenta in addition to the helimagnon
momentum.}
\label{fig:1}
\end{figure}

The spin model described by the Eqs.\ (\ref{eqs:1}) can be understood and
derived as an effective theory that results from an underlying fermionic
action. The technical procedure is to single out the magnetization by either
performing a Hubbard-Stratonovich decoupling of the spin-triplet interaction,
or by constraining the appropriate combination of fermion fields to an
auxiliary composite field whose expectation value is the magnetization
\cite{Hertz_1976, Belitz_Kirkpatrick_Rosch_2006b}. Integrating out the fermions
then yields the spin model, with the coefficients of the LGW theory given in
terms of localized fermionic correlation functions. Conversely, integrating out
the magnetization yields an effective theory of electronic quasi-particles that
interact via an exchange of helimagnons. This effective interaction was derived
and discussed in Ref.\ \onlinecite{Kirkpatrick_Belitz_Saha_2008a}. For small
wave numbers, the leading contribution to the effective potential is
\bse
\label{eqs:8}
\be
V({\bm k},i\Omega;{\bm p}_1,{\bm p}_2) = V_0\,\chi({\bm k},i\Omega)\,
\gamma({\bm k},{\bm p}_1)\, \gamma (-{\bm k},{\bm p}_2).
\label{eq:8a}
\ee
Here $V_0 = \lambda^2 q^2/8(\me^*)^2$ with $\lambda$ the Stoner gap (i.e., the
splitting of the two electron bands that results from the magnetization in a
mean-field approximation) and $\me^*$ the electron effective mass. $\Omega$
denotes a bosonic Matsubara frequency, and
\be
\chi({\bm k},i\Omega) = \frac{1}{2\NF}\, \frac{q^2}{3\kF^2}\,
\frac{1}{\omega_0^2({\bm k}) - (i\Omega)^2}
\label{eq:8b}
\ee
with $\NF$ the density of states at the Fermi surface is the helimagnon
susceptibility. The leading contribution to the vertex function $\gamma$ is
given by
\be
\gamma({\bm k},{\bm p}) = \nu ({\bm k}_{\perp}\cdot{\bm
p}_{\perp})p_{||}/\lambda.
\label{eq:8c}
\ee
\ese
Here $\nu$ is a dimensionless parameter that describes the coupling of the
electrons to the lattice; generically, one expects $\nu = O(1)$
\cite{Kirkpatrick_Belitz_Saha_2008a}.  The effective potential is graphically
depicted in Fig.\ \ref{fig:1}; notice that it depends on the momenta of the
quasiparticles involved in addition to the transferred momentum.

\begin{figure*}[t]
\vskip -0mm
\includegraphics[width=16.0cm]{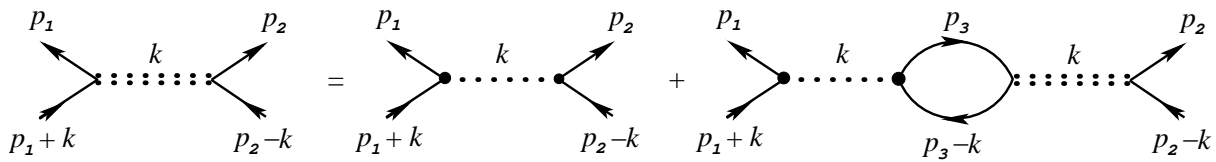}
\caption{Screening of the effective quasiparticle interaction.}
\label{fig:2}
\end{figure*}

Due to the singular nature of the helimagnon susceptibility the effective
interaction is long-ranged, and screening has a qualitative effect. The leading
effect of screening is captured by the usual random-phase approximation (RPA)
\cite{Fetter_Walecka_1971}. The screened interaction is shown in Fig.\
\ref{fig:2}.
If one takes the resulting fermionic theory of quasi-particles interacting via
the screened effective interaction, re-introduces the magnetization, and
integrates out the fermions, one can study the effect of the screening on the
LGW spin model. For our present purposes, the most important effect is a
renormalization of the coupling constant in the crystal-field term
$S_{\text{cf}}$, Eq.\ (\ref{eq:1d}). We find
\be
b \to b - \vert b\vert\,\nu^2\, (\epsilon_{\text{F}}/\lambda)^2
\label{eq:9}
\ee
While the value of the Stoner gap in MnSi is not well known, it is clear from
the small value of the critical temperature that $\epsilon_{\text{F}}/\lambda$
is large compared to unity. An band structure calculation in Ref.\
\onlinecite{Taillefer_Lonzarich_Strange_1986} yielded $\lambda \approx
3,000\,{\text{K}}$, while the Fermi temperature is $T_{\text{F}}\approx
150,000\,{\text{K}}$ \cite{Nakanishi_Yanase_Hasegawa_1980,
parameters_footnote}, for a ratio $\epsilon_{\text{F}}/\lambda \approx 50$. The
effect of the screening is thus large, and the sign of the effect is of
fundamental importance. If $b<0$, then the bare theory yields a helix pinned in
the $(1,1,1)$-direction and the renormalization greatly enhances the
coefficient and hence the pinning strength. If $b>0$, then the bare theory
predicts pinning in the $(1,0,0)$-direction and the renormalization changes
this to produce pinning in the $(1,1,1)$-direction. This is analogous to the
effect of a strong electron-phonon coupling that can trigger a structural phase
transition. The conclusion that the pinning will always be in the
$(1,1,1)$-direction holds for systems where the transition is continuous or
weakly first order. For strongly first order transitions the LGW theory is no
longer controlled, and a gradient-free cubic anisotropy in the action (i.e. a
term proportional to $\sum_i M_i^4$, which we have neglected in Eqs.\
(\ref{eqs:1})) can invalidate this conclusion. These observations are
consistent with experimental results. In MnSi, the transition is continuous or
very weakly first order, and the pinning is in $(1,1,1)$-direction everywhere
in the ordered phase \cite{Pfleiderer_et_al_1997}. In FeGe, the transition is
strongly first order, and the pinning is in $(1,0,0)$-direction close to the
transition, but switches to the $(1,1,1)$ direction at lower temperatures
\cite{Lebeche_Bernhard_Freltoft_1989}.

From the considerations leading to Eqs.\ (\ref{eqs:5}) we see that $\Hcone$ is
proportional to $\sqrt{b}$ while $\Hctwo$ is independent of $b$. The
renormalized theory thus yields the following result for the ratio $\Delta =
\Hcone/\Hctwo$,
\be
\Delta_{\text{theo}} \approx \gso\,\nu\,\epsilon_{\text{F}}/\lambda,
\label{eq:10}
\ee
which replaces Eq.\ (\ref{eq:6}). With the numbers quoted above, and assuming
$\nu \approx 1$, we obtain values for MnSi and FeGe that are in agreement with
the experimental ones given in Eq.\ (\ref{eq:3}) within a factor of $2$.

As a check, we finally discuss the absolute values of $\Hcone$ and $\Hctwo$.
Taking into account the renormalization of $b$, Eq.\ (\ref{eq:9}), $\Hcone$ is
given by
\be
\Hcone \approx \gso\,m_1\,q\,\nu\,(\epsilon_{\text{F}}/\lambda) \sqrt{a/\chi}.
\label{eq:11}
\ee
In the ordered phase of MnSi at ambient pressure, the susceptibility is
observed to be roughly $\chi \approx 6\mu_{\text{B}}^2/\kB\,T_{\text{c}}$
\cite{Pfleiderer_et_al_1997}. This is consistent with theoretical
considerations \cite{Lonzarich_Taillefer_1981}. In a fully renormalized spin
model, the gradient squared term in the action taken at the Fermi length scale
must be roughly equal to the critical temperature, $a\,\kF^2\,m_1^2/2 \approx
T_{\text{c}}$. With these estimates, we obtain from Eq.\ (\ref{eq:11}) $\Hcone
\approx 0.25\,{\text{T}}$, which is the correct order of magnitude. Similarly,
from Eq.\ (\ref{eq:5b}) we obtain $\Hctwo \approx 0.5\,{\text{T}}$, in
agreement with the experimental value.

In summary, we have pointed out that the standard LGW theory for helical
magnets leads to a large discrepancy between the strength of the spin-orbit
interaction in helical magnets as determined from the pitch wave number versus
the ratio $\Hcone/\Hctwo$ of the two critical fields. We have shown that a
renormalization of the theory that results from the screening of the effective
quasiparticle interaction resolves this puzzle and also leads to absolute
values of the critical fields that are in good agreement with the
experimentally observed values.

\acknowledgments

We thank Achim Rosch for comments on the manuscript. This work was supported by
the NSF under grant Nos. DMR-05-29966, DMR-05-30314, DMR-09-01952, and
DMR-09-01907. Part of this work was performed at the Aspen Center for Physics.

\vskip -0mm

\begin{thebibliography}{24}
\expandafter\ifx\csname natexlab\endcsname\relax\def\natexlab#1{#1}\fi
\expandafter\ifx\csname bibnamefont\endcsname\relax
  \def\bibnamefont#1{#1}\fi
\expandafter\ifx\csname bibfnamefont\endcsname\relax
  \def\bibfnamefont#1{#1}\fi
\expandafter\ifx\csname citenamefont\endcsname\relax
  \def\citenamefont#1{#1}\fi
\expandafter\ifx\csname url\endcsname\relax
  \def\url#1{\texttt{#1}}\fi
\expandafter\ifx\csname urlprefix\endcsname\relax\def\urlprefix{URL }\fi
\providecommand{\bibinfo}[2]{#2} \providecommand{\eprint}[2][]{\url{#2}}

\bibitem[{\citenamefont{Ishikawa et~al.}(1976)\citenamefont{Ishikawa, Tajima,
  Bloch, and Roth}}]{Ishikawa_et_al_1976}
\bibinfo{author}{\bibfnamefont{Y.}~\bibnamefont{Ishikawa}},
  \bibinfo{author}{\bibfnamefont{K.}~\bibnamefont{Tajima}},
  \bibinfo{author}{\bibfnamefont{D.}~\bibnamefont{Bloch}}, \bibnamefont{and}
  \bibinfo{author}{\bibfnamefont{M.}~\bibnamefont{Roth}},
  \bibinfo{journal}{Solid State Commun.} \textbf{\bibinfo{volume}{19}},
  \bibinfo{pages}{525} (\bibinfo{year}{1976}).

\bibitem[{\citenamefont{Pfleiderer et~al.}(1997)\citenamefont{Pfleiderer,
  McMullan, Julian, and Lonzarich}}]{Pfleiderer_et_al_1997}
\bibinfo{author}{\bibfnamefont{C.}~\bibnamefont{Pfleiderer}},
  \bibinfo{author}{\bibfnamefont{G.~J.} \bibnamefont{McMullan}},
  \bibinfo{author}{\bibfnamefont{S.~R.} \bibnamefont{Julian}},
  \bibnamefont{and} \bibinfo{author}{\bibfnamefont{G.~G.}
  \bibnamefont{Lonzarich}}, \bibinfo{journal}{Phys. Rev. B}
  \textbf{\bibinfo{volume}{55}}, \bibinfo{pages}{8330} (\bibinfo{year}{1997}).

\bibitem[{\citenamefont{Lebech et~al.}(1989)\citenamefont{Lebech, Bernhard, and
  Freltoft}}]{Lebeche_Bernhard_Freltoft_1989}
\bibinfo{author}{\bibfnamefont{B.}~\bibnamefont{Lebech}},
  \bibinfo{author}{\bibfnamefont{J.}~\bibnamefont{Bernhard}}, \bibnamefont{and}
  \bibinfo{author}{\bibfnamefont{T.}~\bibnamefont{Freltoft}},
  \bibinfo{journal}{J. Phys. Cond. Matt.} \textbf{\bibinfo{volume}{1}},
  \bibinfo{pages}{6105} (\bibinfo{year}{1989}).

\bibitem[{\citenamefont{Dzyaloshinski}(1958)}]{Dzyaloshinski_1958}
\bibinfo{author}{\bibfnamefont{I.~E.} \bibnamefont{Dzyaloshinski}},
  \bibinfo{journal}{J. Phys. Chem. Solids} \textbf{\bibinfo{volume}{4}},
  \bibinfo{pages}{241} (\bibinfo{year}{1958}).

\bibitem[{\citenamefont{Moriya}(1960)}]{Moriya_1960}
\bibinfo{author}{\bibfnamefont{T.}~\bibnamefont{Moriya}},
  \bibinfo{journal}{Phys. Rev.} \textbf{\bibinfo{volume}{120}},
  \bibinfo{pages}{91} (\bibinfo{year}{1960}).

\bibitem[{\citenamefont{Ishikawa et~al.}(1985)\citenamefont{Ishikawa, Noda,
  Uemura, Majhrzah, and Shirane}}]{Ishikawa_et_al_1985}
\bibinfo{author}{\bibfnamefont{Y.}~\bibnamefont{Ishikawa}},
  \bibinfo{author}{\bibfnamefont{Y.}~\bibnamefont{Noda}},
  \bibinfo{author}{\bibfnamefont{Y.~J.} \bibnamefont{Uemura}},
  \bibinfo{author}{\bibfnamefont{C.~F.} \bibnamefont{Majhrzah}},
  \bibnamefont{and} \bibinfo{author}{\bibfnamefont{G.}~\bibnamefont{Shirane}},
  \bibinfo{journal}{Phys. Rev. B} \textbf{\bibinfo{volume}{31}},
  \bibinfo{pages}{5884} (\bibinfo{year}{1985}).

\bibitem[{\citenamefont{Pfleiderer et~al.}(2001)\citenamefont{Pfleiderer,
  Julian, and Lonzarich}}]{Pfleiderer_Julian_Lonzarich_2001}
\bibinfo{author}{\bibfnamefont{C.}~\bibnamefont{Pfleiderer}},
  \bibinfo{author}{\bibfnamefont{S.~R.} \bibnamefont{Julian}},
  \bibnamefont{and} \bibinfo{author}{\bibfnamefont{G.~G.}
  \bibnamefont{Lonzarich}}, \bibinfo{journal}{Nature (London)}
  \textbf{\bibinfo{volume}{414}}, \bibinfo{pages}{427} (\bibinfo{year}{2001}).

\bibitem[{\citenamefont{Bak and Jensen}(1980)}]{Bak_Jensen_1980}
\bibinfo{author}{\bibfnamefont{P.}~\bibnamefont{Bak}} \bibnamefont{and}
  \bibinfo{author}{\bibfnamefont{M.~H.} \bibnamefont{Jensen}},
  \bibinfo{journal}{J. Phys. C} \textbf{\bibinfo{volume}{13}},
  \bibinfo{pages}{L881} (\bibinfo{year}{1980}).

\bibitem[{\citenamefont{Ma}(1976)}]{Ma_1976}
\bibinfo{author}{\bibfnamefont{S.-K.} \bibnamefont{Ma}},
  \emph{\bibinfo{title}{Modern Theory of Critical Phenomena}}
  (\bibinfo{publisher}{Benjamin, Reading, MA}, \bibinfo{year}{1976}).

\bibitem[{par()}]{parameters_footnote}
\bibinfo{note}{Relevant parameter values for MnSi are as follows. The Fermi
  temperature is $T_{\text{F}}\approx 147,000\,{\text K}$
  \cite{Nakanishi_Yanase_Hasegawa_1980}, and the electronic effective mass
  averaged over the Fermi surface is $\me^*\approx 4\me$, with $\me$ the
  free-electron mass \cite{Taillefer_Lonzarich_Strange_1986}. Within a
  nearly-free electron model, this yields a Fermi wave number $\kF \approx
  3.6\,\AA^{-1}$. The exchange splitting or Stoner gap was found to be $\lambda
  \approx 3,300\,{\text K}$ in a band structure calculation
  \cite{Taillefer_Lonzarich_Strange_1986}. We note that the values for
  $T_{\text{F}}$ and $\lambda$ quoted in Ref.\
  \onlinecite{Belitz_Kirkpatrick_Rosch_2006a} were too small by a factor of
  6.24, and the value for $\kF$ was too small by a factor of $\sqrt{6.24}
  \approx 2.5$. In FeGe, the Fermi temperature has been determined only in the
  hexagonal phase, where $T_{\text{F}}\approx 90,000\,{\text K}$
  \cite{Nazareno_Carabelli_Calais_1971}, and the effective electron mass is not
  known.}

\bibitem[{fac()}]{factor_2_footnote}
\bibinfo{note}{Throughout this paper we ignore factors of 2 in our estimates in
  realization of the fact that many parameters that enter the model
  calculations are not known to a better accuracy anwyay.}

\bibitem[{\citenamefont{M{\"u}hlbauer et~al.}(2009)\citenamefont{M{\"u}hlbauer,
  Binz, Jonietz, Pfleiderer, Rosch, Neubauer, Georgii, and
  B{\"o}ni}}]{Muhlbauer_et_al_2009}
\bibinfo{author}{\bibfnamefont{S.}~\bibnamefont{M{\"u}hlbauer}},
  \bibinfo{author}{\bibfnamefont{B.}~\bibnamefont{Binz}},
  \bibinfo{author}{\bibfnamefont{F.}~\bibnamefont{Jonietz}},
  \bibinfo{author}{\bibfnamefont{C.}~\bibnamefont{Pfleiderer}},
  \bibinfo{author}{\bibfnamefont{A.}~\bibnamefont{Rosch}},
  \bibinfo{author}{\bibfnamefont{A.}~\bibnamefont{Neubauer}},
  \bibinfo{author}{\bibfnamefont{R.}~\bibnamefont{Georgii}}, \bibnamefont{and}
  \bibinfo{author}{\bibfnamefont{P.}~\bibnamefont{B{\"o}ni}},
  \bibinfo{journal}{Science} \textbf{\bibinfo{volume}{323}},
  \bibinfo{pages}{915} (\bibinfo{year}{2009}).

\bibitem[{\citenamefont{Plumer and Walker}(1981)}]{Plumer_Walker_1981}
\bibinfo{author}{\bibfnamefont{M.~L.} \bibnamefont{Plumer}} \bibnamefont{and}
  \bibinfo{author}{\bibfnamefont{M.~B.} \bibnamefont{Walker}},
  \bibinfo{journal}{J. Phys. C: Solid State Phys.}
  \textbf{\bibinfo{volume}{14}}, \bibinfo{pages}{4689} (\bibinfo{year}{1981}).

\bibitem[{us_()}]{us_tbp}
\bibinfo{note}{D. Belitz and T.R. Kirkpatrick, unpublished results.}

\bibitem[{\citenamefont{Belitz et~al.}(2006{\natexlab{a}})\citenamefont{Belitz,
  Kirkpatrick, and Rosch}}]{Belitz_Kirkpatrick_Rosch_2006a}
\bibinfo{author}{\bibfnamefont{D.}~\bibnamefont{Belitz}},
  \bibinfo{author}{\bibfnamefont{T.~R.} \bibnamefont{Kirkpatrick}},
  \bibnamefont{and} \bibinfo{author}{\bibfnamefont{A.}~\bibnamefont{Rosch}},
  \bibinfo{journal}{Phys. Rev. B} \textbf{\bibinfo{volume}{73}},
  \bibinfo{pages}{054431} (\bibinfo{year}{2006}{\natexlab{a}}).

\bibitem[{sig()}]{sign_b_footnote}
\bibinfo{note}{In Ref. \protect{\onlinecite{Belitz_Kirkpatrick_Rosch_2006a}}
  the effect of $b\neq 0$ was taken into account qualitatively, and for $b>0$
  only. The result given here follows from analyzing the Gaussian fluctuations
  about a helical state pinned in $(1,1,1)$ direction due to $b<0$.}

\bibitem[{\citenamefont{Hertz}(1976)}]{Hertz_1976}
\bibinfo{author}{\bibfnamefont{J.}~\bibnamefont{Hertz}},
  \bibinfo{journal}{Phys. Rev. B} \textbf{\bibinfo{volume}{14}},
  \bibinfo{pages}{1165} (\bibinfo{year}{1976}).

\bibitem[{\citenamefont{Belitz et~al.}(2006{\natexlab{b}})\citenamefont{Belitz,
  Kirkpatrick, and Rosch}}]{Belitz_Kirkpatrick_Rosch_2006b}
\bibinfo{author}{\bibfnamefont{D.}~\bibnamefont{Belitz}},
  \bibinfo{author}{\bibfnamefont{T.~R.} \bibnamefont{Kirkpatrick}},
  \bibnamefont{and} \bibinfo{author}{\bibfnamefont{A.}~\bibnamefont{Rosch}},
  \bibinfo{journal}{Phys. Rev. B} \textbf{\bibinfo{volume}{74}},
  \bibinfo{pages}{024409} (\bibinfo{year}{2006}{\natexlab{b}}).

\bibitem[{\citenamefont{Kirkpatrick et~al.}(2008)\citenamefont{Kirkpatrick,
  Belitz, and Saha}}]{Kirkpatrick_Belitz_Saha_2008a}
\bibinfo{author}{\bibfnamefont{T.~R.} \bibnamefont{Kirkpatrick}},
  \bibinfo{author}{\bibfnamefont{D.}~\bibnamefont{Belitz}}, \bibnamefont{and}
  \bibinfo{author}{\bibfnamefont{R.}~\bibnamefont{Saha}},
  \bibinfo{journal}{Phys. Rev. B} \textbf{\bibinfo{volume}{78}},
  \bibinfo{pages}{094407} (\bibinfo{year}{2008}).

\bibitem[{\citenamefont{Fetter and Walecka}(1971)}]{Fetter_Walecka_1971}
\bibinfo{author}{\bibfnamefont{A.~L.} \bibnamefont{Fetter}} \bibnamefont{and}
  \bibinfo{author}{\bibfnamefont{J.~D.} \bibnamefont{Walecka}},
  \emph{\bibinfo{title}{Quantum Theory of Many-Particle Systems}}
  (\bibinfo{publisher}{McGraw-Hill, New York}, \bibinfo{year}{1971}).

\bibitem[{\citenamefont{Taillefer et~al.}(1986)\citenamefont{Taillefer,
  Lonzarich, and Strange}}]{Taillefer_Lonzarich_Strange_1986}
\bibinfo{author}{\bibfnamefont{L.}~\bibnamefont{Taillefer}},
  \bibinfo{author}{\bibfnamefont{G.}~\bibnamefont{Lonzarich}},
  \bibnamefont{and} \bibinfo{author}{\bibfnamefont{P.}~\bibnamefont{Strange}},
  \bibinfo{journal}{J. Magn. Magn. Materials} \textbf{\bibinfo{volume}{54-57}},
  \bibinfo{pages}{957} (\bibinfo{year}{1986}).

\bibitem[{\citenamefont{Nakanishi et~al.}(1980)\citenamefont{Nakanishi, Yanase,
  and Hasegawa}}]{Nakanishi_Yanase_Hasegawa_1980}
\bibinfo{author}{\bibfnamefont{O.}~\bibnamefont{Nakanishi}},
  \bibinfo{author}{\bibfnamefont{A.}~\bibnamefont{Yanase}}, \bibnamefont{and}
  \bibinfo{author}{\bibfnamefont{A.}~\bibnamefont{Hasegawa}},
  \bibinfo{journal}{J. Magn. Magn. Materials} \textbf{\bibinfo{volume}{15-18}},
  \bibinfo{pages}{879} (\bibinfo{year}{1980}).

\bibitem[{\citenamefont{Lonzarich and
  Taillefer}(1981)}]{Lonzarich_Taillefer_1981}
\bibinfo{author}{\bibfnamefont{G.~G.} \bibnamefont{Lonzarich}}
  \bibnamefont{and}
  \bibinfo{author}{\bibfnamefont{L.}~\bibnamefont{Taillefer}},
  \bibinfo{journal}{J. Phys. C: Solid State Phys.}
  \textbf{\bibinfo{volume}{18}}, \bibinfo{pages}{4339} (\bibinfo{year}{1981}).

\bibitem[{\citenamefont{Nazareno et~al.}(1971)\citenamefont{Nazareno,
  Carabelli, and Calais}}]{Nazareno_Carabelli_Calais_1971}
\bibinfo{author}{\bibfnamefont{H.}~\bibnamefont{Nazareno}},
  \bibinfo{author}{\bibfnamefont{G.}~\bibnamefont{Carabelli}},
  \bibnamefont{and} \bibinfo{author}{\bibfnamefont{J.-L.}
  \bibnamefont{Calais}}, \bibinfo{journal}{J. Phys. C}
  \textbf{\bibinfo{volume}{4}}, \bibinfo{pages}{2052} (\bibinfo{year}{1971}).

\end{thebibliography}

\end{document}